\documentclass[graybox]{svmult}

\usepackage{newtxtext}
\usepackage[varvw]{newtxmath}
\usepackage{makeidx}
\usepackage{graphicx}
\usepackage{multicol}
\usepackage{footmisc}

\usepackage{type1cm}

\title*{Dark Energy}

\author{Sergio Luigi Cacciatori, Vittorio Gorini and Federico Re}

\institute{Sergio Luigi Cacciatori \at Department of Science and High Technology, Universit\`a dell'Insubria, Via Valleggio 11, Como, 22100, Italy. \\ \email{sergio.cacciatori@uninsubria.it}
\and Vittorio Gorini \at Department of Science and High Technology, Universit\`a dell'Insubria, Via Valleggio 11, Como, 22100, Italy. \\ \email{vittorio.gorini@gmail.com}
\and Federico Re \at Department of Physics Giuseppe Occhialini, Universit\`a di Milano-Bicocca, Piazza della Scienza 3, 20126, Milano, Italy. \\ \email{federico.re@unimib.it}}

\begin{document}

\maketitle

\begin{abstract}
    {Among the great mysteries that physics has not yet solved are undoubtedly those of dark energy and dark matter. In this chapter we deal with the first of them. We will expound in detail the motivations that led to hypothesise the existence of dark energy, the importance of understanding its nature, its main possible theoretical explanations and a list of the many attempts at modelling it.
    We conclude with a description of the most recent and future missions on Earth and in space devised to shed light on this mystery. \\[0.3cm]
    {\it ``This is a preprint of the following chapter: S. L. Cacciatori, V. Gorini, F. Re, Dark Energy, published in New Frontiers in Science in the Era of AI, edited by M. Streit-Bianchi \& V. Gorini, 2024, Springer reproduced with permission of Springer Nature Switzerland AG. The final authenticated version is available online at: \url{http://dx.doi.org/10.1007/978-3-031-61187-2}''.} See \cite{url} for the version of Record.
    }
\end{abstract}

 \vspace{1cm}

\vspace{1cm}

In this and the next chapter, we will discuss the two aspects of what goes under the name of Dark Sector, namely dark energy and dark matter. We start with a short introduction regarding the problematics of both these topics. Then, in the present chapter we will deal with dark energy, in the next with dark matter. However, since these two phenomena are inextricably connected in the unfolding of the evolution of the universe, cross reference to both these topics will often appear in both chapters. Therefore, we warmly invite the readers to read them both.  

\section*{Introduction}
Practically in all fields of science, hitherto more or less well established scientific theories are thrown into question by their incapacity to explain newly found and often mysterious phenomena and discoveries. This stimulates the efforts to formulate new and more encompassing theories, to replace the old ones. In general, this does not mean that the old theories were wrong, but rather that they were approximations valid in certain limiting situations and/or in more restricted contexts. It also often happens that a new theory predicts new and unexpected phenomena to test by experiment or observations. For instance, in physics a paradigmatic example is the Theory of General Relativity of which Newtonian mechanics is the approximation which accurately applies in most situations in which the speeds in play are very small compared to the speed of light and the density of energy is low. An example of a totally new phenomenon predicted by general relativity is the existence of gravitational waves, whose evidence has been recently observed. It is this continuous interplay between theory, experiment and observations which allows us to acquire better and better knowledge of the workings of nature. In this respect, dark energy and dark matter (collectively dubbed the Dark Sector) stand out blatantly. Indeed, these effects are among the greatest mysteries in physics and astrophysics, for each of which no convincing explanation has been found to date. They have been mysteries for a very long time: dark energy for more than a century and dark matter for ninety years. To both effects there applies well the statement that the scientific process starts with the admission of one's ignorance and the precise definition of what the ignorance consists of. For these reasons we will start by defining clearly what we mean by dark matter and dark energy, thus circumscribing our ignorance about such phenomena. First we will debunk some common misconceptions about these topics, in order to make clear what we are talking about and what is the meaning of the words we use. After that we will present the reasons why it is believed that dark matter and dark energy are at the border of our knowledge. In the scientific paradigm, this means that there are one or more empirical observations that we are not able to explain with our existing theories, or that it is not clear how to adapt the theories to explain them. In this and the next chapter we will first discuss the empirical evidences for these two mysterious phenomena, and then list and discuss the attempts that have been made to explain them. We shall see that there are many hypotheses in competition, both for dark matter and dark energy. Finally, we will spend some words on developments that can be expected for the future.\\
The present chapter will be dedicated to dark energy, the next one to dark matter.

\section{Basic concepts}

\subsection{Dark energy is not dark matter}

If the reader has some interest in physics, she/he probably knows that the mass is a form of energy, according to the famous Einstein's formula $E=mc^2$, where $m$ is the mass of any object, $E$ its energy content and $c$ the speed of light in empty space, measuring almost $300,000$ kilometres per second. To get an idea, it takes a bit more than one second for the light originating from the Moon to reach Earth, and about $8$ minutes for the light of the Sun to arrive to us.

However, one should not be misled into believing that dark energy is the energy content of dark matter. They are entirely different objects. They are both hypothetical entities and we believe in their existence because of indirect proofs. The evidences are very different for the two cases. There are many evidences for the existence of dark matter, some of which have an astrophysical nature, while others concern cosmology. On the other hand, we have only two evidences for dark energy, and they come from cosmology alone.


\subsection{The difference between astrophysics and cosmology}

Astrophysics and cosmology are two different, though related fields of science. It is useful to distinguish them for what will follow. Both topics study very large objects, far bigger than anything we are used to deal with in everyday life. However, the cosmological scale is much bigger than the astrophysical one.

Cosmology deals with the study of the universe as a whole. The visible universe is a sphere with a diameter of about $93$ billion light years\footnote{One light year is the distance that light covers in one year. It amounts to about $10^{13}$ kilometres, namely $10,000$ billion kilometres.}, hence this is the typical field of application of cosmology. Cosmological speculations are also concerned with what happens beyond this sphere, thus encompassing even bigger regions of space. \\
On the contrary, astrophysics is interested in limited objects, such as planets, planetary systems, stars, star clusters, galaxies, galaxy clusters and superclusters. These objects are typically held together by gravity, so that they rarely exceed $100$ million light years in extension. \\
For comparison, if the size of the visible universe were reduced to that of a large city, the diameter of a rocky planet like the Earth would be smaller than an atomic nucleus and a galaxy cluster would have a diameter of the order of a puma.

In both cosmology and astrophysics, gravity, which is described by Einstein's Field Equations of general relativity, is the dominant interaction and manifests itself as an attractive force. However, cosmology is also subject to some kind of "exotic" behaviour, as we shall see.

\subsection{General Relativity and cosmological models}\label{GRC}

To date, general relativity is the best theory we have for the description of spacetime and the action of gravity \cite{Einstein popular, Abc}, \cite{Weinberg book}. The earlier paradigm, due to Isaac Newton, considered space and time as absolute entities, completely unrelated to the motion of objects. They form an absolute, homogeneous and isotropic grid, within which bodies move under the action of various forces. According to this paradigm, all forces are generated by bodies. A body which is not acted upon by any force moves inside absolute space along a straight line at constant speed\footnote{The Principle of Inertia.}. On the other hand, if some force acts on the body, the latter will be subject to an acceleration proportional to the acting force and inversely proportional to the body's mass. In this context, gravity is a particular force, which is always attractive: two bodies attract each other with a force which is directly proportional to the product of their masses and inversely proportional to the square of their distance. This is Newton's Universal Law of Gravitation\footnote{For example, among several other things, this law explains with great accuracy the motion of comets and the motion of planets along elliptical orbits around the Sun}.

Newton's paradigm was overturned by Einstein in $1905$ \cite{Einstein 1905} through his formulation of the theory of Special Relativity and, in $1915$ \cite{Einstein 1915, Einstein 1916}, by generalising the latter to incorporate gravity, with the formulation of the theory of general relativity. Special relativity debunked the Newtonian postulate of space and time having each a separate absolute nature, by showing that they are relative entities which depend on the state of motion of the observer who performs the measurements. Space and time get mixed up between themselves by changing by one observer to another in relative motion\footnote{A well known example is the famous adventure of two twins, one of whom stays at home, whereas the other leaves for a space travel. When he comes back to meet his brother he is now younger than him. Time has flown differently for the two.}. In this way, space and time meld into a unique four-dimensional continuum dubbed spacetime. However, in special relativity spacetime keeps a flat nature, namely it obeys the rules of Euclidean geometry. \\
On the other hand, in general relativity the effect of gravity is to bend spacetime by making it curved. Namely, masses, mass currents (moving masses), any form of energy, energy flows and pressure, do curve, compress, stretch and deform the fabric of spacetime in a dynamical way. In such a picture, a body which is subject to gravity alone glides unhindered and without friction along the curved spacetime continuum: the body is falling freely. In other words, whereas electric and magnetic fields still act as \emph{bona fide} forces on charged particles, according to general relativity gravity is not a force anymore. It is replaced by free, unhindered motion along curved spacetime. \\
Curvature embodies gravity. With a metaphor, we may say that, before Einstein, space and time were just the stage of a theatre, in which matter, energy, forces and fields acted as characters of the drama. With general relativity the stage itself comes to life and joins the other characters in the dance.

The general relativistic paradigm was soon applied to cosmology, to calculate the geometry and the evolution of the whole universe. Though already in the late tens of last century Einstein himself realised that his equations could be applied to this purpose, the first concrete model of the evolution of the universe was elaborated in $1922$ by the Russian mathematician Alexander Friedmann \cite{Friedman 1922, Friedman 1924}. He observed that, at sufficiently large scales (several megaparsec\footnote{One megaparsec (Mpc) equals one million parsec and one parsec (pc) is equal to $3.26$ light years (lyr).}) inhomogeneities in the distribution of matter in space would be smoothed out. Therefore, to a first approximation, the distribution of matter and energy of the universe could be taken as homogeneous and isotropic\footnote{These terms mean that all places and, respectively, directions are equivalent. Therefore, we call "inhomogeneous" a distribution that is denser somewhere and less dense somewhere else. The Earth is, for instance, a clear inhomogeneity in the matter distribution, since it is much denser than the space around it. The same for the Sun and for the whole Galaxy. However, the billions of galaxies we see are distributed in an homogeneous way, so that at very large scales the universe is approximately homogeneous.}. Hence, the universe's curved geometry, which is generated by this distribution and which represents the gravitational field, would share the same properties. This meant that the Friedmann universe is described by a few parameters: the (current) density of matter, the (current) density of radiation, and the (current) curvature of space, which could be either positive, or zero, or negative\footnote{\label{curvature}To understand the meaning of positive and negative curvature, consider the simple example of a two-dimensional surface in our familiar three-dimensional Euclidean space. At some given point P of this surface the curvature is positive if a sufficiently small neighbourhood of P on the surface lies entirely on one side of the plane tangent to the surface at the given point. On the other hand if, no matter how small the neighborhood at P we choose, part of it will lie on one side and part on the other side of the tangent plane at the surface at P, then the curvature at the point is negative. Since the central point of a saddle has this property, a point of negative curvature is called a saddle point.}. Contrary to what was widely believed at that time, such a universe cannot be static, but it must either expand or contract. Hence, since matter-energy attracts itself gravitationally, if at present the universe were still, in the past it must have expanded until its current size, and it would contract in the future. Otherwise, if we were to require that the universe would never contract and collapse, then at the present epoch it would be expanding and the initial propulsion had to be strong enough as to avoid any future collapse. The situation is similar to someone standing in the surface of the Moon and throwing a rock vertically in mid air\footnote{We envisage the experiment taking place on the Moon and not on Earth because the Moon has no atmosphere thus avoiding complications due to air friction.}. If the initial speed of the rock is not too large, the latter will decelerate up to a certain height, then invert the direction of its motion and fall back to the ground. However, if the initial upward speed of the rock were equal to or larger than a critical speed (called escape speed\footnote{The escape speed for the Moon is $2.4$ kilometres per second.}), the rock, though decelerating, would overcome the gravitational attraction of the celestial body and fly off to space without ever falling back. In this metaphor, the rock represents the universe. The highest point above the surface of the Moon achieved by the rock represents the maximum extension attained by the universe, just before it starts to contract. The gravitational attraction exerted by the Moon corresponds to the gravity of all the matter-energy content of the universe. The rock fallen back on the ground corresponds to the collapsed universe shrunk to a final state of infinite density. The speed of the rising rock symbolises the universe expansion rate. Finally, the case when the initial speed of the rock is equal to, or larger than, the escape speed, represents the situation when the initial expansion rate of the universe is so large that the latter will keep expanding forever, though at a gradually reducing speed. This metaphor is nice because the Friedmann equations for the universe expansion are mathematically identical to the dynamical equations for the rock.

Einstein was aware that his theory required a dynamical universe. However, he thought this to be an absurd conclusion, as he was convinced that the universe was static. Indeed, that was the general mindset of those years. Therefore, in order to cure the problem, Einstein added a suitable extra term to his equations, multiplied by a constant which he called the cosmological constant and denoted by $\Lambda$. The extra term in Einstein's equations acts as a gravitational repulsion and, by choosing for $\Lambda$ a suitable value, it compensates exactly the expansion rate, thus giving rise to a static universe. On the other hand, in $1929$, Edwin Hubble \cite{Hubble 1929}, using the largest telescope of the time, discovered that distant galaxies recede from us with a speed proportional to their distance, thus confirming after all that the universe is expanding and validating Friedmann's model. Unfortunately, Friedmann didn't live long enough to witness the triumph of his model, since he had passed away two years earlier. The cosmological constant wasn't necessary anymore and was thus dropped, and Einstein himself acknowledged his mistake.

Today, the Friedmann model is referred to as the FLRW geometry, after the scientists A. Friedmann, G. Lema\^itre, H.P. Robertson and A.G. Walker, who independently developed the model in the $1920$s and $1930$s \cite{Friedman 1922, Friedman 1924}, \cite{Lemaitre 1931a, Lemaitre 1931b}, \cite{Robertson 1935, Robertson 1936a, Robertson 1936b}, \cite{Walker 1937}. Tracing back to the past, the model predicts an initial state of incredibly high mass-energy density and temperature, as was first proposed in $1927$ by the Belgian Roman Catholic priest and physicist Georges Lema\^itre. In other words, the universe is assumed to have originated from an initial so called Big Bang, a term introduced by the British astronomer Fred Hoyle \cite{Kragh}, followed by a continuous expansion and cooling, up to the present epoch. A broad range of observed phenomena confirm the Big Bang theory. Among them, in particular, the Cosmic Microwave Background Radiation (CMB) discovered in $1964$ by Arno Penzias and Robert Wilson \cite{Penzias Wilson 1965}, namely the left over electromagnetic radiation originated when the universe was approximately $380,000$ years old, the time at which photons ceased to be energetic enough to be able to ionise neutral atoms and therefore started to travel freely through space, resulting in the decoupling of matter and radiation. The FLRW geometry is an exact solution of the Einstein field equations of general relativity describing a homogeneous, isotropic, expanding (or otherwise contracting) universe. It depends on two parameters, a scale function $a(t)$ giving the rate of expansion (or contraction) as a function of time, and a number $k$ which takes either one of the three values $1$, $0$ or $-1$ depending on whether the curvature of space is positive, zero or negative. Depending on the matter-energy content of the universe, the scale factor $a(t)$ may characterise a never ending expansion or an expansion followed by a contraction.

\section{The evidence for dark energy}\label{DE evidence}

In order to understand the meaning of the expressions "dark energy" and "dark matter", we must consider and interpret the observational clues pointing to the existence of such objects, which will also justify the names given to them.

In this chapter we deal with dark energy and defer to the next chapter the treatment of dark matter. Until quite recently, there was just one observational clue pointing to the existence of the effect that we call dark energy, and it does not concern astrophysical objects. Rather it has a purely cosmological nature: it is the \emph{redshift of far away type Ia supernovae}. On the other hand, in the last few years, a second important phenomenon has proved to be relevant in indicating the action of dark energy, in the evolution of the early universe and in the formation of structures. It is the phenomenon of Baryon Acoustic Oscillations (BAO). We will be concerned with supernovae first and treat BAO later.

We then start by explaining the meaning of redshift. Consider first the term \emph{frequency shift}. If we listen to the sound of a motorcycle or of an ambulance approaching to us, the pitch of the sound emitted by these vehicles, and heard by us, is higher than the one we feel after, having passed by, the vehicles recedes from us. What is the reason for this phenomenon? Sound is produced by pressure waves propagating through the air. The wavelength $\lambda$ of a wave is the distance between two successive wave crests (in the case of sound, it is the distance between two successive planes of highest air pressure). The frequency $\nu$ of the wave is equal to the speed v of sound propagation divided by the wavelength: $\nu={\rm v}/\lambda$.\footnote{The speed of sound in air at sea level and at a temperature of $20$° C is about $343$ metres per second. It is the difference between this speed and the enormous speed of light which is responsible for the fact that we hear thunder quite later after having seen the lightning.} The pitch of a sound wave in inversely proportional to its wavelength and proportional to its frequency: shorter wavelengths, hence higher frequencies, correspond to higher pitches. Longer wavelengths (lower frequencies) to lower pitches. When the motorcycle or the ambulance is approaching us the sound waves it emits are compressed to shorter wavelengths, hence we experience the higher pitch. On the other hand, when the vehicle recedes from us, the waves are stretched to longer wavelengths and we experience a lower pitch.

This is the so-called Doppler effect, after the name of its discoverer, the Austrian mathematician and physicist Christian A. Doppler. It applies to all kinds of wave propagation, including electromagnetic waves, hence, in particular, light: borrowing a terminology taken from the rainbow order, when a luminous object is receding from us, and hence the frequency of light we receive  from it is shifted to lower values, we speak of a \emph{redshift}. Conversely, if the source of light is moving towards us, the frequency of the light we perceive is shifted to higher values and we speak of a \emph{blueshift}. The amount of frequency shift is expressed by a number, conventionally denoted by $z$. This number is respectively positive or negative depending on whether the source of light is redshifted or blueshifted and the absolute value of $z$ gives a precise estimate of the speed at which the source is receding from or, respectively approaching, the observer ($z=0$ corresponds to a source which is still with respect to the observer\footnote{The exact formula for the frequency shift is $1+z=\gamma(1+{\rm v}\cos\theta /c)$, where $\gamma=1/\sqrt{1-{\rm v}^2/c^2}$ and $\theta$ is the angle between the direction of motion of the source of light and the direction of emission in the observer's frame. $\theta=0$ corresponds to the source receding directly away from the observer (maximum redshift), $\theta=\pi$ to the source moving directly towards the observer (maximum blueshift). For $\theta=\pi/2$ we have $1+z=1/\sqrt{1-{\rm v}^2/c^2}$, the transverse redshift.}).

It was precisely thanks to the observation of this frequency shift that Hubble discovered that the universe is expanding. Indeed, he concluded that distant galaxies are receding from us because he observed that the spectral light emitted by the elements composing the atmospheres of their stars was redshifted. And, in addition, he also found that the speed of any receding galaxy is proportional to its distance from us. In other words, he established the relation v$=H_0 d$, where v and $d$ are respectively the receding speed and the distance of the given galaxy and $H_0$ is a constant which has gone down to history with the name of \emph{the Hubble parameter}\footnote{"Late universe" measurements of the $H_0$ have yielded values of approximately $73$ (km/s)/Mpc. Whereas "early universe" techniques based on measurements of the CMB agree on a value near to $67.7$ (km/s)/Mpc. This discrepancy is called the Hubble tension. The cause for this discrepancy is unknown, though several proposals have been put forward to solve it. We come back to this problem later on (see subsection \ref{noise}).}. Accordingly, the proportionality relation is called the \emph{Hubble law}. This law has been a spectacular confirmation of Friedmann's model of an expanding universe, based on the equations of general relativity: each galaxy is approximately still in space, but it is space itself which is expanding, quite like a huge elastic sheet that is subject to uniform stretching in all directions.

But how was Hubble able to evaluate the distance of the galaxies he observed? This is not a trivial problem. On a telescope, a far star is no more than a brilliant point, and a galaxy looks like a luminous nebula. The color of the spectral lines gives us the speed, but what about the distance? To measure the distance of a galaxy, Hubble exploited the properties of a special kind of stars, the so-called Cepheid variables, discovered in $1908$ by the American astronomer Henrietta S. Leavitt while studying variable stars in the Magellanic Clouds \cite{Leavitt 1908}. The term Cepheid originates from the first example of a star of this kind, found in the constellation Cepheus, in $1784$. The Cepheid variables are stars pulsating in luminosity and for which a well defined relation exists between the period of pulsation and the absolute luminosity of the star. Therefore, upon comparing the apparent luminosity of one such star with its absolute luminosity determined by the observation of its period of variability, one can accurately determine its distance\footnote{The \emph{intrinsic luminosity} $L$ of a source of light is the total amount of energy radiated by the source per unit time. It is usually measured in watts. At a distance $r$ form the source, this energy gets spread uniformly over the surface of a sphere, with area $4\pi r^2$. The \emph{apparent brightness} $b$ of the source, measured by an observer located at the given distance $r$, is the amount of light from the source that crosses the unit area per unit time. It is thus given by $b=L/4\pi r^2$, and measured in watts per square metre (W/m$^2$). \\
If we have two sources having the same intrinsic luminosity and located at distances respectively $r_1$ and $r_2$ from the observer, the ratio of their apparent brightness $b_1$ and $b_2$ is thus $b_1/b_2=(r_2/r_1)^2$.}. \\
In astronomy, a \emph{standard candle} is a source of light that has a known absolute luminosity. Cepheid variables, which are common in all galaxies, are standard candles that can be used to measure distances up to $30$ Mpc. Beyond such distances, their light becomes too weak to be observed. Hence, to evaluate distances farther off, more powerful standard candles were needed. In the late eighties such a powerful standard candle become available: the \emph{type Ia supernovae}.

We explain this term. Stars are essentially huge thermonuclear bombs whose energy comes from the fusion of hydrogen into helium and heavier elements in their interior. However, the fusion energy which would make the star explode is balanced by the force of gravity which would tend to make the star collapse onto itself. On the other hand, when the nuclear fuel is exhausted, this balance is broken, and the star begins to contract. For not too big stars such as our sun, this process leads to the formation of a fluffy outer layer called a red giant that eventually blows away, while the rest contracts into a very dense, hot and stable core called a \emph{white dwarf}. Instead, for much bigger stars, typically having a mass of about $8$ solar masses or more, the process is catastrophic and the star explodes, suddenly becoming extremely luminous and scattering around a large amount of debris that flies off in all directions at enormous speeds. In general, the central core collapses into a neutron star or a black hole, though sometimes it happens that the whole mass of the stars flies away in fragments. \\
Such an explosion is called a \emph{supernova}. It is one of the most energetic phenomena in the universe, which can reach luminosities several billion times that of the Sun. The absolute luminosity of a supernova depends on the mass of the dying star. \\
However, there exists a particular class of stars that happen to explode when they all attain essentially the same mass, therefore giving rise to supernovae all with practically the same luminosities. The phenomenon takes place whenever a white dwarf is in a close orbit around a companion star that is slowly losing matter from its outer atmosphere, falling by gravity on the dwarf, which therefore keeps increasing its mass. Then, as soon as the mass of the dwarf reaches a critical limit, which is of the order of $1.4$ times the mass of the Sun, gravity suddenly overcomes the balancing effect of internal pressure and the dwarf explodes into a supernova whose luminosity is thus essentially always the same. These supernovae are termed of \emph{type Ia} and can be employed as standard candles\footnote{Actually, it has turned out that the intrinsic luminosity of type Ia supernovae is not strictly uniform. However, the differences can be compared and catalogised, allowing these supernovae to be at least standardisable candles. \\
Type Ia supernovae can be used to measure distances up to about $1000$ Mpc, hence less than $z=1$. On the other hand, astronomers have recently found that X-ray and ultraviolet luminosities of quasars are tightly correlated, making them suitable to be used as standard candles up at least to $z\sim7$.}.

\subsection{The cosmological constant strikes again}
\label{121}

During the years from $1994$ to $1998$, two groups of researchers, led respectively by Brian Schmidt and Adam Riess \cite{Riess 1998}, and by Saul Perlmutter \cite{Perlmutter 1999}, took up the job of employing a large sample of type Ia supernovae at vastly different redshifts, with the purpose of confirming with precision Hubble's law and studying how the expansion of the universe was decelerating over time, according to Friedmann's model. Indeed, the light coming from more distant supernovae would take a longer time to reach Earth, thus providing us with information on the rate of expansion at ever earlier times for increasing redshifts. \\
However, to the great surprise of both teams, the results of the measurements indicated that, in the past, the rate of expansion was slower than in the present epoch. In other words, it turned out that, instead of decelerating, the expansion of the universe \emph{was accelerating}. This was a quite unexpected and revolutionary conclusion. It was, however, readily accepted by the community of astrophysicists, since the results of the two teams agreed, though they worked independently and had used mostly different sets of supernovae. Hence, after all, Einstein hadn't been entirely wrong, and some cosmological constant could account for the puzzle. However, it has to be remarked that thirty years earlier, in the late sixties, Yakov Zel'dovich had realised that the energy carried by the vacuum fluctuations of quantum fields would act as a cosmological constant \cite{Zeldovich 1967a, Zeldovich 1967b}. However, original estimates of this contribution yielded a phantasmagorical value of about $120$ orders of magnitude larger than what could reasonably be expected, and the idea was left aside. We will pick up this problem again in the next Section. \\
Going back to the metaphor of the rock, the cosmological constant can be pictured as a tractor beam from some flying saucer that acts on the rock by pulling it vertically upwards with a constant force $\Lambda$, whereas gravity attracts the rock downwards. $\Lambda$ is a constant force, whereas gravity gets weaker as the rock climbs upwards. In Einstein's idea, the height of the rock above the ground was such that the force of gravity and $\Lambda$ were in perfect balance, so that the rock would stand still in mid air. On the contrary, the observations of Schmidt, Riess, Perlmutter and collaborators indicated that, at the present height of the rock, gravity is weaker than the pull $\Lambda$ of the tractor beam, so that the rock keeps accelerating upwards ever faster.

On the basis of observations, the Friedmann model is described by four numbers: the mean curvature of space, which appears to be zero; the present expansion rate $H_0$, and the components that fill the universe, namely matter and radiation. The energy density of radiation is very small and can be neglected. From these parameters, the model foresees other cosmological quantities, among which a quantity denoted by $q_0$ and called the deceleration parameter, as it would give a measure of the rate at which the expansion of the universe would be slowing down. \\
Instead, Schmidt, Riess and Perlmutter found that the expansion is accelerating and the value they determined for the deceleration parameter was therefore negative: $q_0\cong-0.53$. Adding to the model a cosmological constant $\Lambda$ to provide for the observed acceleration, this implied for $\Lambda$ the value $\Lambda\cong1.1\times10^{-52}$ m$^{-2}$. It is a very tiny value indeed. Such a repulsion would be too small to be perceptible in the everyday life, or even by its effects on the orbits of the planets, or on the behaviour of the whole galaxy. But once the whole volume of the universe is considered, its repulsive contribution becomes relevant. \\
Formally, the introduction of a cosmological constant is equivalent to introduce a new energy component that fills homogeneously the whole universe, besides matter and light. However, while matter and light exert a positive pressure, giving an attractive gravity, such new component gives rise to a \emph{negative} pressure, resulting in a repulsion. The energy density of the cosmological constant amounts to $c^4\Lambda/8\pi G$, where $G$ is the gravitational constant appearing in Newton's universal law of gravitation\footnote{Newton's universal law of gravitation states that two masses $m_1$ and $m_2$, located at distance $r$ between them, attract each other gravitationally with a force given by the formula $F=Gm_1 m_2/r^2$, where $G$ is called the gravitational constant.}, and the negative pressure that it exerts is the density with the minus sign. In order to fit the observed values of the deceleration parameter, the energy associated to the cosmological constant would constitute $68.5$\% of the total energy-matter content of the universe, the remaining $31.5$\% being due to the normal (baryonic\footnote{The term ``baryonic'' originates from the Greek ``$\beta\alpha\rho\acute\upsilon\varsigma$'', meaning ``heavy''. Normal matter (making up stars, planets, and ourselves) is composed by atoms and molecules. In turn, an atom is made by a nucleus, composed by protons and neutrons, orbited by electrons. Proton and neutrons are much heavier than electrons, hence are cumulatively called baryons.}) matter and the dark matter, with radiation contributing only with $0.00001$\%.

Of course, any form of energy exerting a negative pressure could explain the accelerated expansion, not just a cosmological constant, or a cosmological constant alone, or even may be something entirely different in nature. \\
To summarise, in order to explain the present day accelerated expansion of the universe, we need some new exotic form of energy. This energy is invisible and to date unknown and utterly mysterious. It has been named "\emph{dark energy}" \cite{Huterer Turner 1998, Perlmutter Turner White 1999}. It was discovered $26$ years ago and its nature is today still as mysterious as it was then. It may fill up space uniformly everywhere and be constant in time, or it may even vary in space and time. Who knows?

The accelerated expansion of the universe was discovered by looking at the redshift of type Ia supernovae. However, as anticipated earlier, there exists nowadays another potentially powerful method to help uncover the nature of dark energy. It is based on the study of the propagation of baryon acoustic oscillations (BAO). Let us explain what this means. \\
During the hot plasma era of the early universe, in which baryons (protons, neutrons and light nuclei), electrons and photons were continuously mutually interacting, local pressure inhomogeneities produced sound waves propagating through the plasma outwards from the original over-dense regions of ordinary matter and dark matter. Whereas dark matter, which interacts only gravitationally, stayed put in the original over-dense regions, these acoustic waves of ordinary matter and photons propagated outwards from these regions in spherical shells up to the time of the decoupling of matter from radiation, about $378,000$ years after the Big Bang, when the pressure dropped quite suddenly to almost zero and the shells froze in place. \\
The maximum distance a sound wave can travel in the primordial plasma before the plasma cools to the point when neutral atoms form and the plasma wave stops expanding, is called the sound horizon and measures about $490,000$ light years in the expanded universe today. It provides a \emph{standard ruler} for the later formation of structures like galaxies and galaxy clusters. Indeed, the baryons and dark matter formed configurations that included over-densities both at the original sites of the anisotropies and on the corresponding shells at the sound horizons of these anisotropies. Such anisotropies eventually became the overlapping ripples in matter densities that would later form galaxies and galaxy clusters. These over-density configurations frozen in place at the time of decoupling are still printed and visible in the CMB. Then, the measurements of the pattern of distances between densities of oscillations in the CMB can be compared to the growth of these distances over space and time, as these densities eventually give rise to galaxy and galaxy clusters. In this way, one can gather information about how the rate of expansion of the universe varies in time under competition between gravity and the repulsive effect of dark energy. {In \cite{Haridasu 2017} it is shown that while measures on type Ia supernovae provide a marginal proof of the accelerated expansion (with confidence $3\sigma$, see section \ref{noise} for an explanation of this notation), the method of oscillations in CMB provides a stronger evidence (with confidence $5.38\sigma$).}

\begin{figure}
\label{Figure1}
	\begin{minipage}[b]{1.0\linewidth}
		\centering
		\includegraphics[width=10 cm]{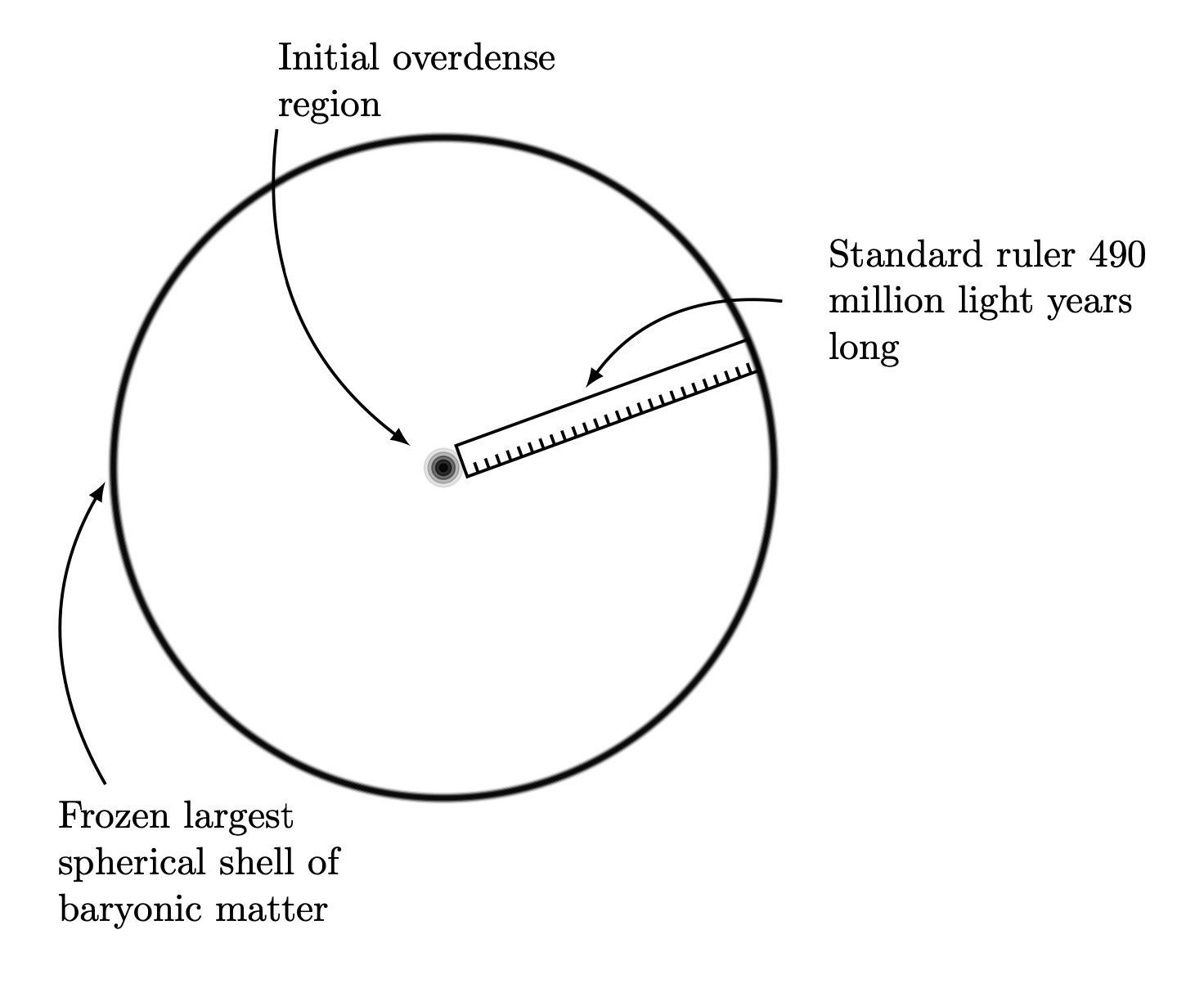}
		\caption{Pictorial representation of an overdense region (centre of the disc) and the circle representing the farthest traveled frozen sound wave (sound horizon).}
	\end{minipage}
\end{figure}

There are at present three main instruments performing major BAO surveys, which should allow the reconstruction of cosmic evolution at ever earlier eras across the universe. The Dark Energy Spectroscopic Instrument (DESI) mounted on a telescope at Kitt Peak National Observatory in Arizona. It is designed to collect optical spectra for more than $40$ million galaxies, quasars and stars, that will allow to construct a three-dimensional map extending from the nearby objects back to a time when the universe was about a quarter of its present age. \\
The Prime Focus Spectrograph (PFS), an instrument on the $8.2$ metre Subaru Telescope on Manua Kea, Hawaii, that will start operating this year and extending the results of DESI at even greater distances. \\
And the European Space Agency's Euclid, which was launched on July 1, $2023$, that will also contribute its own survey of galaxy evolution to the BAO catalog. In addition, Euclid will also add to the investigation of the nature of dark energy through the method of gravitational lensing, the phenomenon which exploits the magnifying effect of the general relativistic bending of the light coming from distant sources by foreground objects such as galaxies and galaxy clusters. This technique will itself allow the study of the growth of galaxy clustering over time under the competition between the gravitational attraction of matter and the repulsive effect of dark energy. \\
If dark energy happens not to be constant but varying in space and time, DESI, PFS and Euclid might be able to tell.

These and some future instruments devised for the study of the properties and nature of dark energy will be more thoroughly discussed in the concluding section 1.4.

\section{Hypotheses on the nature of dark energy}

We have seen that, formally, the introduction of a cosmological constant is equivalent to the requirement of the existence of a new, exotic form of energy component of the universe, that has been called dark energy and that exerts a negative pressure, thus driving the expansion of the universe to accelerate. To date, the nature of dark energy is unknown. However, some hypotheses can be advanced as to the nature of this remarkable phenomenon. We will expound them here. \\
In the sequel we will use the terms "dark energy" and "cosmological constant" interchangeably.

\subsection{Residual curvature}

The first hypothesis is that the cosmological constant has a purely geometrical origin, namely a nature entirely independent of the presence or not of matter and energy. In order to clarify this point, we may imagine how how a totally empty spacetime would appear. To begin with, first consider just space.

It is reasonable to believe that in empty space, namely in the absence of bodies or of any form of energy like for instance electric or magnetic fields, no position and no particular direction is privileged. All points of space are equivalent (space is homogeneous) and all directions originating from any given space point are equivalent (space is isotropic). How do homogeneous and isotropic spaces look like? To get the idea, start first with a two-dimensional space, namely a surface. \\
To an ant walking on a perfectly flat, uniformly coloured and infinitely extended table all points and all directions will appear equivalent. But the ant would have the same feeling if it were walking on the surface of a perfect sphere: all points and all directions on the sphere are likewise equivalent. Both the flat table (an infinite plane) and the spherical surface are two-dimensional homogeneous and isotropic spaces. The difference is that the plane has \emph{zero curvature} (it is flat), whereas the sphere is curved: it has a \emph{constant positive curvature}\footnote{The equation of the sphere is $x^2+y^2+z^2=R^2$ and its (positive) curvature is $K=1/R^2$. As the radius $R$ of the sphere grows the curvature decreases and in the limit when $R$ becomes infinite ($R\rightarrow\infty$) the sphere turns into the flat Euclidean plane $\mathbb{R}^2$ of zero curvature.}. \\
Do there exist homogeneous and isotropic two-dimensional spaces having \emph{constant negative curvature}? All points of any such surface should be equivalent saddle points (see note \ref{curvature}) with equivalent directions. Now, in our familiar flat three-dimensional Euclidean space $\mathbb{R}^3$ it is easy to construct two dimensional surfaces whose points are all saddle points\footnote{\label{negative}A simple example of one such surface is the one-sheeted hyperboloid, with equation $x^2+y^2-z^2=R^2$.} \cite{Henle book}. However, in all such examples the negative curvature varies from point to point (in the example of note \ref{negative}, it decreases as the absolute value $|z|$ of $z$ increases) and, at any point, isotropy is lost. But now try conceiving in $\mathbb{R}^3$ a two-dimensional homogeneous and isotropic surface whose points are all saddle points, namely a surface with constant negative curvature ($K=-1/R^2$) and that is isotropic at all points. You will not succeed! Indeed, you will realise immediately that it is impossible to mentally conceive such a surface. The reason is that a surface of this kind does indeed exist, but it cannot be embedded in $\mathbb{R}^3$. It can only be embedded in the flat Euclidean four-dimensional space $\mathbb{R}^4$.

Moving now from two-dimensional surfaces to three-dimensional spaces, the requirement of homogeneity and isotropy is again realised in three ways: the zero curvature case which is our familiar flat Euclidean space $\mathbb{R}^3$; the three-dimensional sphere\footnote{The three-dimensional sphere can be embedded in $\mathbb{R}^4$ and its equations is $x^2+y^2+z^2+w^2=R^2$.}; and the three-dimensional isotropic space with constant negative curvature. \\
The Einstein's equations of general relativity imply that the presence of matter and energy curve and deform not just space but spacetime as well. If, by removing all matter and energy, spacetime remains curved, its curvature will be called \emph{residual curvature}. It is the curvature that survives, after all other possible bending and deforming causes have been removed. \\
Note that in an empty spacetime the homogeneity property extends to time: all time instants are equivalent among themselves in a totally empty universe. We will speak of homogeneous time, hence of homogeneous spacetime. Whereas it would make no sense to talk of spacetime isotropy: isotropy is a property of space alone.

Just as there exist homogeneous and isotropic three-dimensional spaces, empty homogeneous spacetimes with isotropic space slices exist as well. Their residual curvature is precisely their (geometrical) cosmological constant. They are respectively called de Sitter space if their cosmological constant (= residual curvature) is positive and anti-de Sitter if it is negative. The case of zero cosmological constant corresponds to the flat Minkowski spacetime of Special Relativity\footnote{The spacetime of the Friedmann model is homogeneous and isotropic but not empty. Its curvature is constant in space but varies in time.},\footnote{As stated earlier, the cosmological constant $\Lambda$ can be seen as a measure of curvature, as well as a measure of a density of dark energy. Precisely, if $\rho$ is the dark energy density, from Einstein's equations one finds $\Lambda=\frac{8\pi G}{c^4}\rho$, where $c$ is the speed of light, $\rho$ is measured as energy per unit volume and $\Lambda$ has the dimension of inverse square length, which is indeed a measure of curvature.}. All these spacetimes are homogeneous and spatially isotropic, each of them being characterised by its particular value of the cosmological constant or, equivalently, of its curvature.

There is no particular motivation why one or another of the values of the cosmological constant should be privileged. Hence, from this point of view, there is neither an explanation for the value measured, nor a reason to believe that such value should be more or less odd or preferred compared to any other. It would be quite astonishing if this value were exactly zero. \\
The independence of this value from the presence or not of matter and/or energy would raise the (geometrical) cosmological constant to the state of a universal constant just as such is the speed of light in vacuum (relative to any local coordinate system). However, as we will now discuss, quantum effects suggest other possibilities.

\subsection{Quantum fluctuations}

The preceding hypothesis leads to a more complicated conclusion if one takes into account the quantum nature of things. The classical description of an electric field or of a magnetic field is simple. The field is there if its intensity is different from zero and it disappears if the intensity is zero. To assert that the field is not there or that it is zero everywhere are equivalent statements in a classical theory. But the true nature of fields is quantum and things are different: either the field does not exist at all or, if it exists, it manifests itself even when its value is zero. This happens because the quantum world is restless, hence its vacuum state doesn't have anything to do with the idea we usually have of it. \\
A manifestation of this fact is already present in ordinary quantum mechanics: it is the well known Heisenberg uncertainty principle. The motion of a particle of mass $m$, such as an electron, a proton or a grain of sand, is quantified by its momentum $\vec{p}=m\vec{v}$,\footnote{we employ the usual convention of indicating vectors with boldface letters. A vector $\vec{V}$ is characterised by its magnitude $V=|\vec{V}|$ and by the direction in which it is pointing. The position vector $\vec{X}$ is characterised by its components $x, y$ and $z$ relative to some coordinate system in $\mathbb{R}^3$. Same for the components ${\rm v}_x, {\rm v}_y$ and ${\rm v}_z$ of the speed vector $\vec{V}$, etc.} the product of its mass times its velocity. Classically, the motion of the particle is characterised by the values that its position in space $\vec{x}$ and its momentum $\vec{p}$ take at any given instant. Quantum mechanics asserts that, in any given space direction, such as for instance the x axis, it is not in any way possible to simultaneously associate with an arbitrary precision the value of the component $x$ of the particle position and the corresponding value of the component $p_x$ of its momentum. This is a consequence of the fact that any physical property of the particle (or, for that matter, of any physical body), such as position, momentum, energy etc., is in general intrinsically uncertain, to a larger or lesser extent depending on the particular state of the system under consideration. \\
In the case in question of the coordinate $x$ and the component $p_x$ of the momentum of the particle, this property is expressed by the relation $\Delta p_x \Delta x\geq h/4\pi$, meaning that the uncertainty $\Delta x$ affecting the position multiplied by the corresponding uncertainty affecting the momentum cannot be less than $h/4\pi$, where $h$ is the so called Planck constant whose value is $h\cong 6.626196\times 10^{-27}$ erg s (erg times second) \cite{Heisenberg 1927}. This is a very small number indeed, $0.000\dots01$ with 26 zeros between the dot and the $1$. So small that it does not affect in any appreciable way macroscopic bodies (except when cooperative effects such as, for instance, superfluidity or superconductivity take place), such as grains of sand for example. But it is an important number in the subatomic world. Let us see why. \\
Recall that an electron has a mass of $0.911\times10^{-27}$ g (about one billionth of a billionth of a billionth grams). And that the size of a typical atom is $10^{-8}\sim10^{-7}$cm. Therefore, electrons orbiting the nucleus of an atom are confined within such typical distances. Hence, by the uncertainty principle $\Delta x\times \Delta p\sim h/(4\pi)$, the uncertainty in the speed of the electron is of the order of $3000$ Km per second $=3\times10^8$ cm/s, about one hundredth the speed of light.\footnote{There are several ways to establish the dimension of atoms. For example, an estimation of the atomic size is based on the assumption that the radius of an atom is half the distance between adjacent atoms in a solid.}
We can interpret these calculations as follows. A proton which captures an electron by its electrical attraction to form a hydrogen atom, confines the electron into a very small region, of the order of a tenth of a nanometer\footnote{1 nanometer$=10^{-9}$ m$=10^{-7}$ cm}. Thus, by virtue of the uncertainty principle, the electron is indeed forced to move inside that region at a speed not less than the estimated value of about one hundredth the speed of light. That's the quantum restlessness. \\
According to classical theory, an accelerated electrically charged body radiates in the form of electromagnetic waves. Therefore, the negatively charged electron which, orbiting the proton, is accelerated, should loose energy by radiation and fall onto the proton. But the electron obeys the laws of quantum mechanics and the uncertainty principle prevents this collapse from taking place.

The uncertainty on the momentum and on the position implies also an uncertainty in the energy attributable to a given particle, as the energy of a particle depends in general on both the momentum (kinetic energy) and on the position (potential energy). Starting from the above-mentioned Heisenberg principle, it is not difficult to prove that the uncertainty on the energy that one can attribute to a particle times the time necessary to measure it, cannot be much less than $h$. In this version, the uncertainty principle has a very important consequence: the principle of energy conservation can be violated by a quantity $\Delta E$ for a time interval $\Delta t$, provided the latter is small enough, namely such that the product $\Delta E \Delta t$ would not be much less than $h$. Since $h$ is so small, this corresponds to the fact that any violation of energy conservation takes place in a time so short that none of such violations can be measured. These phenomena, leading to a non observable violation of energy conservation, are called \emph{quantum fluctuations} \cite{Mandelstam Tamm 1945}.

We are now ready to discuss the manner in which the quantum vacuum is deeply different from the classical one and how this could have to do with dark energy. \\
Recall that all fields like the electromagnetic field, are characterised by wave properties so that we can always imagine them composed by a superposition of several monochromatic waves of different wavelengths, or equivalently, of different frequencies\footnote{Recall the relation $\lambda={\rm v}/\nu$, where $\lambda$, v and $\nu$ are, respectively, the wavelength, the speed and the frequency of the given monochromatic wave.}. Such decomposition is known as Fourier decomposition and every component with given frequency is said to be a mode of the field. Everybody knows the example of how a glass prism decomposes the solar light into different colours, each corresponding to a given monochromatic wave. The same happens in the rainbow, where the prism (or the crystals of a chandelier) is replaced by droplets of water in which the light behind us gets reflected to reappear to us divided in its colours. \\
Well then, not only the electromagnetic field, but all fields can be thought as a sum of monochromatic waves, each with its own intensity, called amplitude of the wave, depending on the situation. In classical theory, the empty field corresponds to the situation in which the amplitude of every monochromatic wave is zero. Hence, the whole field is zero and it's like not being there. What changes for the corresponding quantum field?

The change is a consequence of the uncertainty principle. Consider the case in which the quantum field is zero so that we are tempted to say that there is no field at all. We expected then that all Fourier modes of the field are absent. We could ascertain this by attempting to measure any one mode. However, such a mode is characterised by its frequency, which is the inverse of the period, the time necessary to perform an oscillation. Therefore, we cannot measure the presence or not of the mode in a time shorter than such a duration, which represents thus the minimum measurement time for the detection of such a mode. \\
But the mode carries an energy proportional to its intensity. Hence the quantum fluctuations can take place in such time lapse with an energy $\Delta E\cong h/\Delta t\cong h\nu$. Therefore, in quantum theory the classical vacuum is replaced by a continuous seething of quantum fluctuations. These fluctuations can all disappear only if the field is simply non existing. \\
For example, the quantum fluctuations  of the electromagnetic field would all disappear only if electromagnetic phenomena would simply not be there in the universe: a universe without electromagnetism. A universe with just strong (namely, nuclear), weak and gravitational interactions, but no charged particles, no light, no X-rays, no electromagnetic interactions at all! How would such a universe look like? Would you be able to conceive it?

Another, perhaps more appealing way of looking at quantum fluctuations, is the following. \\
All elementary particles, electrons, neutrinos, photons, quarks etc., are described by excitations of the corresponding quantum fields that describe them. Photons are the excitations (quanta) of the electromagnetic field, electrons and positrons of the Dirac field, and so on. Now, consider the earlier relation $\Delta E\cong h/\Delta t\cong h\nu$ for the minimum time needed to detect and measure a given mode of a given quantum field. It is simply an expression of the well known quantum mechanical wave-particle duality. Indeed, it tells us that exciting a wave of frequency $\nu$ corresponds to the creation of a field particle of energy $h\nu$, which lives a time of the order $\Delta t$ before disappearing. This means that quantum fluctuations can be interpreted as continuous creations and subsequent annihilations of particles, taking place uniformly everywhere in space and time. Any such particle pops out from nothing with an energy $\Delta E$ for a very brief time $\Delta t$ such that $\Delta E \Delta t\cong h$, after which it disappears. \\
This is the reason why particles created by quantum fluctuations are called \emph{virtual}. In particular, since for a given virtual particle $\Delta E$ cannot be smaller than its rest energy, $\Delta t$ will be of the order of $h/mc^2$ or less, which is in general a very tiny time indeed. For example, in the case of an electron, the smallest charged particle\footnote{Since electric charge is not subject to uncertainty, virtual electrons are actually excited in pairs with virtual positrons.}, one finds that $\Delta t$ is at most of the order of $10^{-10}$ s.

Note that it cannot happen for a field to be at times classical and at other times quantum. Classical fields are just approximations, valid in specific situations in which quantum effects are negligible. But the true nature of fields is quantum and we cannot neglect quantum fluctuations in the energy budget. As stated, these fluctuations are homogeneous, they do not depend on the points in space, nor on time, hence they contribute with an everywhere uniform energy density. To calculate this density, one must consider how many modes are present per unit volume for each frequency, then sum over all frequencies. \\
This is the most delicate part of the calculation, since it requires to know which are the admissible frequencies. To this purpose, recall that a frequency of oscillation is inversely proportional to the wavelength: higher frequencies correspond to smaller wavelengths. We may think that there may exist a limit to the smallest possible frequencies, due to the possible finite extension of the universe. However, the universe is anyway so large that there would correspond such large wavelengths and hence such small frequencies that they would not appreciably contribute to the counting of the total energy density\footnote{A photon with a wavelength as large as the observable universe would have an energy of about $10^{-33}$ electronvolts (one millionth of a billionth of a billionth of a billionth eV) to be compared to the mass of an electron whose energy equivalent is $511,000$ electronvolts. See footnote \ref{eV}.}. \\
Instead, it is interesting to know if there is a minimum value for the wavelengths under consideration. This takes place at the scale at which it is impossible to maintain the belief that our intuition of a spacetime continuum would keep its validity. Such scale is dictated by the mixing of quantum and gravitational effects and is estimated to be of the order of Planck length\footnote{Planck length is given by the formula $l_P=\sqrt{hG/2\pi c^3}$, where $G$ is the Newton's universal constant of gravitation, $h$ is Planck constant and $c$ is the speed of light.}, which amounts to about $1.6\times10^{-33}$ cm. Using the corresponding frequency as the maximum admissible one, and taking into account all the fields that come into play in the standard model of elementary particles, it is finally possible to estimate the total energy contribution due to the fluctuations.

The value obtained in this way is not the exact one but rather an estimation as it depends in detail on the precise value of the largest admissible frequency, of which at present we do not have a sure knowledge. Thus it is rather an estimation of the order of magnitude. \\
Now, the energy density corresponding to the value experimentally observed to date of the cosmological constant is of the order of $3.35$ GeV/m$^3$ \footnote{\label{eV} 1 GeV$=10^9$ eV. eV is the symbol used for electronvolt, which is the energy an electron acquires upon being accelerated through a potential difference of one volt. 1 eV$=1.602\times10^{-12}$ erg.} \cite{Planck 2018}. While, on the other hand, the aforementioned theoretical estimation leads to values that are between $10^{120}$ and $10^{123}$ larger. A huge discrepancy!

However, this does not necessarily mean that the calculation is wrong. Indeed, we should also consider the contribution of the residual curvature previously discussed. The two effects are not mutually exclusive. It may well be that adding them up they would give the observed value. This tells us that the observed value of the cosmological constant is by no means incompatible with the theoretical considerations made so far, at least until we might be able to find a separate way to measure the quantum fluctuations and the residual curvature. \\
However, if this were truly the situation, we would be confronted with another mystery that we could not fail to believe would require a very specific explanation. Indeed, if it were true that the energy due to the quantum fluctuations is $10^{120}$ times larger than the observed value, it would mean that the residual curvature compensates it almost entirely, letting only the observed value to survive. This would mean that the two values are equal for at least 120 figures, a coincidence unmatched by any other phenomenon and not a very credible one unless justified by some very deep significance.

An interesting thing that one may think should be taken into account is that the vacuum energy of fermions (particles whose spin is semi-integer, such as electrons or neutrinos) is, for each degree of freedom, equal to that of bosons (particles with integer spin, such as photons) but with opposite sign\footnote{Reasoning in classical terms, it turns out that, in general, each elementary particle rotates about an intrinsic axis with a certain angular velocity, somewhat like a spinning top. The amount of such rotation is called the \emph{spin} of the particle and, for any given particle, it equals the product of $h/(2\pi)$ times either an integer or a half integer. For example an electron has spin $\frac{1}{2}\frac{h}{2\pi}$, a photon has spin $\frac{h}{2\pi}$, etc. For simplicity, the factor $\frac{h}{2\pi}$ is understood and one says that an electron has spin $\frac{1}{2}$, a photon has spin 1, etc. Particles with integer spin are called \emph{bosons}, particles with half integer spin \emph{fermions}.}. Hence, if bosons and fermions had the same number of degrees of freedom, these energies would perfectly neutralise themselves and only the geometrical contribution of the residual curvature would survive, without the need to compensate anything. \\
However, this is not the case in the standard model: the number of degrees of freedom of the fermionic fields is different from that of the bosonic fields. A possible solution could be provided by a theory based on some kind of symmetry that would set into direct correspondence bosonic fields with fermionic fields, in such a way that the number of degrees of freedom would be the same for both. \\
One such symmetry is called \emph{supersymmetry} and it could in principle solve the dilemma of the cosmological constant. However, there is no proof of the validity of such hypothesis, which would imply the existence of new particles, the so called superparticles, which so far have never been found. Other possible explanations of such coincidence (such as the anthropic principle and quantum cosmology) have not had to date better success than the supersymmetric theory, hence we omit any further discussion.

\subsection{Backreaction from inhomogeneities}

A possible alternative explanation of the accelerated expansion of the universe (as well as some effects normally ascribed to dark matter) are based on the idea that it is the a priori assumption of homogeneity and isotropy of the universe that requires a cosmological constant \cite{Buchert 2018}. On the other hand, it is obvious to everybody that the universe is far from being truly homogeneous and isotropic. \\
For example, strictly speaking space homogeneity corresponds to the fact that, at a given instant, all measurable physical quantities of the universe should be the same at all points in space. This is patently false: for example, at "small scales" we see that stars are interspersed with empty space (or anyway containing much less dense matter, such as gas and dust). Therefore, we can state that, where stars are there, the density of matter is much higher than in their surrounding space. However, if we now start considering a portion of galaxy containing many stars, we will realise that they are approximately equally distributed in that region, provided the latter is not too small nor too large. We may then think it reasonable to replace the precise local description of matter with its average value, given by the total mass contained in this portion of space divided by its volume. Of course, this average description will be effective at the scale at which we have calculated the average value, whereas at smaller scales we would realise that, upon changing our position from one place to another, things would change. Then, we may expect to be able to take into account these changes in terms of small deviations from the average. \\
It's a bit like observing a glass full of perfectly still water. The water will appear to us as a perfectly homogeneous substance, taking up the entire space inside the glass, and describing it in this way works very well at the scale of the glass. However, if we had at our disposal an extremely powerful microscope we would realise that this is not how things stand, the glass would appear to us full of molecules made up of atoms, whose masses are concentrated in the nuclei, the diameters of which are a hundred thousands times smaller than those of the atoms themselves, and therefore occupying a space a million of a billion times smaller. Hence, the space inside the glass is mostly empty, with extremely dense masses concentrated here and there\footnote{The density of any atomic nucleus, from hydrogen to uranium, is of the order of a billion tons per cubic centimetre, which means that a glass full of such matter would weight about one hundred billion tons. Neutron stars have densities of this order, and even higher in their core.}. In addition, each water molecule keeps bouncing off other molecules and moving far and wide inside the glass. However, this is not relevant as long as we are only interested in the scale of the glass and in the macroscopic properties of the water, such as its density and temperature.

We are led to assert that we can work similarly with our universe. The average values that we obtain depend on the scales at which we calculate them. The average mass density of the whole galaxy does not coincide with the one that we would obtain in a portion at the centre of the galaxy or in a portion at the periphery. At the border of the galaxy the gas of stars is considerably more rarefied than inside. \\
On larger scales we will observe the average density to decrease further, since our galaxy is surrounded by space filled by rarefied intergalactic gas. Hence, outside the galaxy the density quickly drops to very low values, until we meet the next galaxies. Galaxies form clusters, which in turn group themselves into superclusters interspersed among almost empty spaces. \\
These different manifestations of grouping of matter at various scales are called \emph{structures}. Structures are the things that determine anisotropies and inhomogeneities at the various scales.

It is here that the primary criticism to the standard cosmological model ($\Lambda$CDM), which is based on the validity of the FLRW equations, comes into play. This model relies on the assumption of space isotropy and homogeneity for what concerns \emph{all} physical quantities. Let us see a bit more in detail what this assumption means and on which hypotheses it relies on. \\
First of all, homogeneity and isotropy are assumed on the scale of the entire universe at a fixed time. Our observations are not made on the entire universe, but only on a portion of the universe, the portion that we see and that we can see (the observable universe). Thus, one assumes that average values calculated on the observable portion of the universe are the same as for the whole universe. \\
However, since we have seen that the calculated averages are sensitive to the scale, this assumption may be a source of error, for example should new structures exist at scales scales larger than the observable universe. Note also that the assertion "at a fixed time" is anything but trivial. It means to admit that it is possible to synchronise clocks globally in such a way as to be able to define a universally valid time (a \emph{global cosmological time}), a highly non obvious fact in general relativity. It is however a crucial assumption in any cosmological model, and we are not interested in discussing it further. What is important to say is that it is only the homogeneity assumption that allows us to directly bind such cosmological time to the proper time\footnote{The proper time of something or of someone is the time measured by a clock that is at rest with respect to that something or that someone.} of the constituents of the universe, which otherwise would neither be unique nor synchronisable\footnote{Let us go back to the example of the glass full of water. We do not consider it strange to associate a common time to all points of the water, by associating to each point a synchronisable clock. And we can obviously associate such time with the proper time of these clocks. But we again peer into our microscope we will see that this time does not have anything to do with the ones of the single molecules that instead move chaotically inside the glass, in such a way that we find it impossible to define a coordinate system relative to which every molecule is still. We have succeeded in the operation of synchronisation within the macroscopic glass because we have "homogenised" the water by replacing the physically relevant quantities with their corresponding average values.}.

To replace the universe by its average is therefore an operation necessary for the scope of being able to identify (or at least to set into correspondence) the cosmological time with the proper time of local observers, such as we ourselves are. By all means the FLRW model abides by this principle, but somehow in a too restrictive way. It assumes that the spatial curvature of the universe (taken to be small or zero) satisfies a kind of conservation law: its product by the square of the expansion parameter is a constant. This requirement is certainly valid for a space exactly homogeneous and isotropic, but not necessarily (and in general not at all) for a space that has been rendered homogeneous and isotropic by a space average. The motivation of this fact, identified by Thomas Buchert et al., is the following (see \cite{Buchert 2018} and references therein). \\
Suppose that we observe the total distribution of matter in its exact time evolution (namely without having calculated its average). If, at each instant of the evolution, we calculate the average distribution of matter, what we obtain is the average of the time evolution of the distribution of matter. Conversely, we may first calculate the average of the matter distribution, and then study the time evolution of such average. Now the question is: do the two procedures lead to the same result? In other words: solving the Einstein equations for the exact distribution of matter and then taking the average would give the same result as first taking the average of the distribution, inserting it in the Einstein equations and then solving them? \\
Intuitively, we are tempted to answer positively since the total mass is conserved and hence the density of matter depends only on the total volume and not on the detail of how matter has or hasn't accumulated into structures. If we believe that through the Einstein equations geometry should evolve in the same way as matter, then this answer should not surprise us. This is what happens in the FLRW model thanks to the geometric assumption that we have mentioned. However, the fact is that while the law of mass conservation is a consequence of the laws of dynamics, the abovementioned "law of conservation of curvature" is an additional assumption but not a consequence of any fundamental law. \emph{We have good reasons to believe it to be valid in the FLRW model but not in the general, since in general relativity the geometry participates fully to the dynamics.} \\
Then, if we remove such forced request we find that the answer changes. The formation of structures is due to the local collapse of matter caused by the gravitational attraction leading to local increases in density surrounded by regions of void. Within these inhomogeneities it happens that the spatial curvature becomes positive in the regions of high matter density and negative in the voids. This favours the expansion of the void regions with an effect much greater with respect to the case in which one had replaced the inhomogeneities with the average matter density and the average curvature. \\
In order to clarify the meaning of this fact, let us take a portion of the universe and a system of local comoving coordinates, namely such that the parts of the universe of this portion have constant space coordinates during the considered time interval. In particular, the portion of universe in question is chosen in such a way that the amount of matter contained in it, and also containing the void regions, is always the same: its mass doesn't change. Then, since the voids expand more rapidly than predicted by the equations when to all quantities we associate their average values, we conclude that the average value of the evolution of the system is different from the time evolution of the average. This is the effect of the structures on the local curvature. This effect is called \emph{backreaction from inhomogeneities}.

Therefore, we conclude that the FLRW equations, which describe the evolution of a spatially perfectly homogeneous and isotropic system, are not the correct ones for the purpose of describing the evolution of a non homogeneous universe. If we include the effect of the inhomogeneities, these equations get modified by suitable additional terms (called source terms) that correctly determine the average of the time evolution of the universe. \\
It is then important to clarify that, on the basis of what we have seen, the average values of the various physical quantities depend on the scale at which they are calculated. Therefore, we can only expect that such additional terms would depend on the details of the scales of observation.

It is known, mainly from the deep field observations of the Hubble Space Telescope (HST) that, in the course of structure formation, galaxies, galaxy clusters and superclusters have progressively arranged themselves in relatively thin filaments and walls that are at present up to a few million light years in extension, and that these structures form the boundaries of enormous bubbles that are practically void of matter. As we have seen, in the course of the expansion the voids, on account of their negative curvature, grow in size at a much faster rate than what is predicted by the FLRW model. And, according to Buchert et al., it is precisely this curvature effect, and not a cosmological constant, which is responsible for the accelerated expansion of the universe. \\
On account of this expansion, filaments and walls will eventually dissolve into isolated clusters moving away from each other at ever increasing speeds. And, according to Buchert's program, it is precisely the addition of the correct source terms to the FLRW equations which would give rise to the accelerated expansion.

Ultimately, then, the transition from decelerated to accelerated expansion, that took place when the universe was a few billion years old , would have taken place as soon as the structures started to organise themselves into filaments and walls, thereby allowing the formation of large voids bounded by them and letting the negative curvature take the lead. \\
In other words, Buchert assumes that there is no intrinsic cosmological constant, either due to a residual curvature, or to the effect of quantum fluctuations, or to the combined effect of both these contributions. Instead, that a "cosmological constant" builds up itself progressively by the action of the negative curvature of the ever increasing voids.

An alternative approach to a "dark energy" effect arising from inhomogeneities has been proposed by David Wiltshire, Asta Heinesen and collaborators. These researchers believe that the accelerated expansion of the universe is simply not there, that it is an artefact of the fact that we observe from inside a large structure, a very long filament that includes the Virgo supercluster, to which our galaxy belongs. In other words, according to these authors it is the effect of the gravitational attraction of all matter contained into this mammoth structure that simulates an accelerated expansion when we look at the recession of the far away type Ia supernovae. \\
Without entering into the details of their work, we can explain the effect by the working of a simple model. Precisely, consider the particular case of the general relativistic Lema\^itre-Tolman-Bondi (LTB) \cite{Lemaitre 1933, Tolman 1934, Bondi 1947} isotropic evolution of a ball of pressureless fluid (dust), in which the dust is expanding outside a sphere $R(t)$ and contracting towards the centre inside. The expansion at values of the radial coordinate $r$ larger than $R(t)$ is of course decelerated because the gravitational attraction of the dust slows down its rate. Now consider an observer freely falling with the dust, and located at a distance $r_0(t)$ inside the ball, and observing the dust outside, in the radial direction (see Fig.2). Since the experimenter is falling towards the centre of the ball at an accelerated rate, if he observes in the direction of his shorter distance from $R(t)$, depending on the rate of his fall he may well see the dust outside the ball moving away from him at an accelerated rate. This effect is of course artificial, since the observer is not still but falling inwards towards the centre at an accelerated rate. \\

\begin{figure}
\label{Figure2}
	\begin{minipage}[b]{1.0\linewidth}
		\centering
		\includegraphics[width=10 cm]{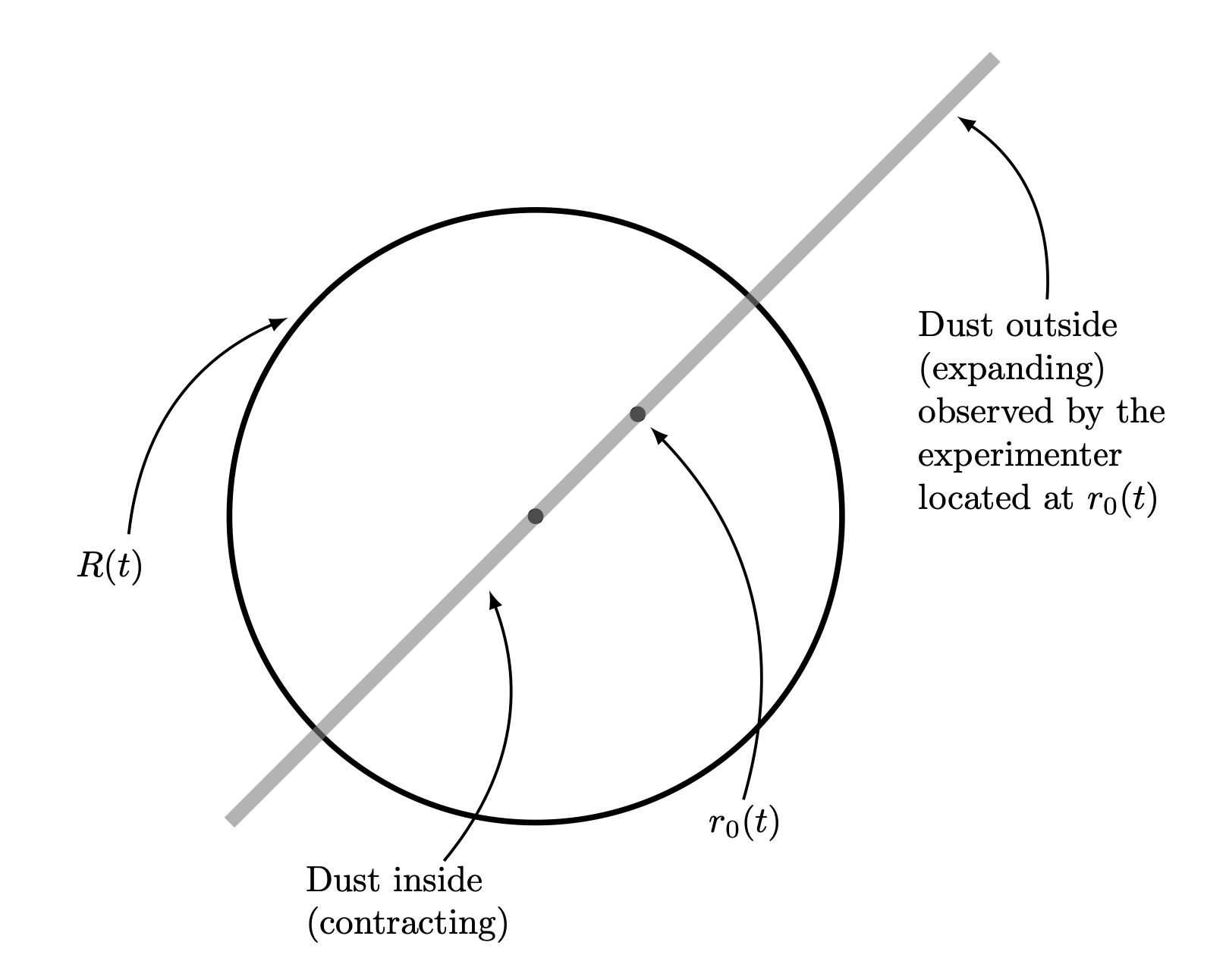}
		\caption{}
	\end{minipage}
\end{figure}

This model is not a particularly good one as it is asymmetrical: if the experimenter observes in the opposite direction, he may well even see the outside dust falling radially towards him. It gives however the idea of the argument put forward by Wiltshire \cite{Wiltshire 2009} and Heinesen \cite{Asta 2022} that the accelerated expansion of the universe may be an artefact of the gravitational effect of the structure inside which are conducting the observations.

The possible explanation of the universe proposed by Buchert et al. is conceptually very enticing since, when applied to smaller scales, it may even solve some issues connected with dark matter, without the need to introduce the latter, to modify general relativity, and to introduce a cosmological constant from the outset. However, we should remark that both Buchert's, and Wiltshire and Heinesen's approaches would leave open the problem of a cosmological constant as a consequence of the vacuum quantum energy. Why would the cosmological constant not appear at all in spite of the great contribution predicted by quantum theory? Perhaps because of perfect compensation by the residual curvature? If this were the case, we believe that it could hardly be a coincidence and there should be some fundamental justification for such a perfect compensation. But this would be another mystery, quite independent of the one at the basis of the accelerated expansion.

\subsection{Noise from inhomogeneities}
\label{noise}

There is another very recent model, inspired by the idea of analysing the role of inhomogeneities. The Buchert's method, expounded in the preceding paragraph, tackles the issue of inhomogeneities by averaging over suitable scales. This leads to corrections to the FLRW equations which depend on the scale, and which become progressively smaller as the scale increases. In a very recent paper by Andrea Lapi et al. \cite{Lapi 2023}, the authors propose an alternative analysis for the largest inhomogeneities scales, in which they replace the averaging method by a stochastic analysis. For "largest scales" we mean the following. \\
It is known that over distances beyond about $100$ Mpc the universe appears to be homogeneous and isotropic. On the other hand, over distances less than about $10$ Mpc it appears strongly inhomogeneous. Therefore, the idea is to study the evolution equations of the universe at an intermediate scale of the order of a few tens of Mpc. An exact analysis of the equations is impossible, since they assume an extremely complicated form: structures appear as a kind of web composed by walls and filaments entangled to form nodes where the density of galaxy clusters is higher. These webs envelop large regions of void. The proposed method of analysis consists in abandoning any attempt at an exact analytic description while instead replacing the FLRW equations by \emph{stochastic equations}. \\
The stochastic calculus is a very efficient technique, employed in very complex systems for which it is not possible to determine in an exact manner the data which define the system under study, but they can be instead characterised with a certain probability. An important example is the classical Brownian motion, through which Einstein was able to prove the existence of atoms. Very small pollen particles placed in a glass of water move with small zigzag jerks in arbitrary directions. This directionally random motion of the pollen particles is due to their collisions with the water molecules, that however we do not see, hence about whose motion we do not have any information. Therefore, we can make the hypothesis that at each instant, with a certain probability, a molecule would hit a particle, sending it jerking in some direction from some position. We do not know this probability a priori, but it will become a parameter to be measured once we have in this way determined the resulting equations for the motion of the pollen particles and compared the solutions with the observations.

A similar technique is applied to the universe at the largest scales, by describing the influence of the inhomogeneities by means of the addition of a source of probabilistic disturbance in the FLRW equations, called \emph{stochastic noise} (a little ado for everything)\footnote{In addition to this, one has also to modify the method of calculation. Indeed, the stochastic movements are still continuous but their speeds are not, as they change by discontinuous steps. The appropriate method of calculation for this situation is called Stratonovich's stochastic calculus \cite{Schilling Partzsch book}.}. We then say that at such scales the universe evolves according to a stochastic dynamics. The result is that the dynamics favours the expansion of the regions of space with less density of matter (the void spaces) in a way entirely similar to the one predicted by the method of backreaction, but with a larger intensity at these scales. \\
In this way, the method reproduces the accelerated expansion of the universe without the need to introduce a cosmological constant, and therefore without dark energy. The effect normally attributed to dark energy would therefore arise as a consequence of the presence of a stochastic noise, traditionally indicated by the Greek letter $\eta$, and therefore called by the authors the $\eta$CDM model in place of the $\Lambda$CDM. If this were all, one might think that there would not be a great advantage in replacing the $\Lambda$CDM model with the new one, based on more complicated methods of calculation. \\
After all, the cosmological constant could have a very simple explanation in terms of a residual curvature and, moreover, the mystery of how the enormous effects of the vacuum fluctuations would disappear would still be there even in the new model. But $\eta$CDM solves two important problems which are present in $\Lambda$CDM: the problem of the \emph{Hubble tension} \cite{Hu Wang 2023} and the \emph{coincidence} problem \cite{Velten vom Marttens Zimdahl 2014}. Let us see what these problems are about and how the $\eta$CDM model solves them.

The Hubble tension concerns the fact that two distinct ways of measuring the Hubble expansion parameter $H_0$, about which we spoke in Section \ref{DE evidence}, give different results. The first method is a direct measurement and it consists in observing the redshift of type Ia supernovae (the standard candles). This is also called the Late Time Measurement. It gives for $H_0$ a value of about $73$ km/s/Mpc. The second method consists instead in an indirect measurement, based on the observation of the spectrum of the CMB (Subsection \ref{GRC}). Observing this cosmic background, we can study its component frequencies, both in neighbouring regions, corresponding to recent times, as well as at extremely large distances, for which the light has taken billions of years to reach us. The latter frequencies therefore correspond to a primordial universe (baby universe). For this reason this is called Early Time Measurement. The point is that in the meantime the universe has expanded, thereby "stretching" the wavelength of the electromagnetic field. This stretching depends on the expansion of the universe, hence on $H_0$, the value of which we can find from the measure of the stretching. It is in this sense that we are dealing with an indirect measurement. In order to obtain in this way the value of $H_0$ we must use the equations of motion, that in turn depend on the model. In this way, the $\Lambda$CDM model gives the value $H_0\cong67$ km/s/Mpc. This discrepancy is the problem. \\
The reason is that when we perform a measurement, we also estimate the error within which the result can be considered reasonably correct. This error is determined by calculating the so called standard deviation $\sigma$. Once this is known, we write the result in the form $x\pm n\sigma$, where $n$ is an integer. This means that the true value of the quantity being measured falls into the interval $[x-n\sigma; x+n\sigma]$ with a probability that depends on $n$. For $n=1$ the probability is $68$\%, for $n=2$ it is $95$\% etc. The errors on the two preceding measurements (having assumed $\Lambda$CDM to be valid) do not coincide up to $n=4$, corresponding to a probability of $99.99$\%. This means that the intervals around the first and the second value do not overlap, and that therefore the results of the two measurements of $H_0$ do not coincide with a probability of $99.99$\%. This suggests that there is something wrong in the $\Lambda$CDM model. \\
On the other hand, if we perform the calculations of the indirect measurement with $\eta$CDM in place of $\Lambda$CDM, one obtains a value for $H_0$ which is compatible within $1\sigma$ with the one obtained by means of the direct measurement, hence the discrepancy disappears.

Instead, the cosmic coincidence problem concerns the value of the cosmological constant (namely, of the corresponding energy density). The point is that the universe has acquired its present state through an expansion which has lasted several billion years, starting from a situation of enormous matter-energy density. During the expansion, this density kept decreasing as matter got ever more diluted in space, having attained today values comparable with those of the cosmological constant. On the other hand, since the energy density of the cosmological constant does not change in time, while the matter-energy density keeps on decreasing due to the expansion, how come that the two densities happen to have acquired comparable values right at the time of our existence on Earth, when it hasn't been so in the far past, nor it will be so in the future? Is there a reason for such a coincidence? Or is it by pure chance that the two values happen to be comparable right at the present epoch? \\
From the point of view of scientific analysis, coincidences are always suspicious and, before accepting them, it is preferable to look for explanations that would render coincidences very probable. Natural solutions to the problem of cosmic coincidence are difficult to envisage. For example, as a possible solution the \emph{anthropic principle} has been called for (Brandon Carter, 1974 \cite{Carter 1974}). Quoting Steven Weinberg \cite{Weinberg 1989}: "Briefly stated, the anthropic principle has is that the world is the way it is, at least in part, because otherwise there would be no one to ask why it is the way it is." There are various reformulations of the principle but none of them really appears to represent an explanation, but rather a reformulation of the problem. \\
In the $\eta$CDM model, instead, the solution appears in an extremely natural way: the stochastic equations allow to predict the value of the density of matter, the deceleration parameter, etc., for arbitrary large times. The result is that the values predicted for infinite times (asymptotic values) are very close to the ones measured today. These values are predicted to change very slowly in time once they get close to the asymptotic value. This means that they are close to such value since quite a long time, which explains why the present one is a natural value. The probability to find a given value for a given quantity is high when the value is taken on for a long time. Therefore, it seems that the $\eta$CDM model is a very good candidate to replace $\Lambda$CDM.

One may wonder if the extension of such analysis to scales smaller than the largest ones may bring to some novelty even in the matter of dark matter.

\subsection{Alternative theories}

There exist many more alternative proposals for solving the problem of dark energy, together with several other problems, among which the attempts to reconcile gravity theories with quantum theory. Since it is impossible to describe all these proposals in a limited space, we confine ourselves to give them a quick overview, without any claim to illustrate them, but only with the purpose of providing the interested reader with the key words useful for personal inquiry.

The common idea of almost all these theories is to add extra matter, fields or other ingredients that would in some way replace the cosmological constant. First of all there are three main lines \cite{Clifton 2012, CANTATA}. \\
\begin{enumerate}
    \item \underline{Scalar-tensor theories}. The main idea is that of introducing a scalar field which couples to gravity but not necessarily to matter fields. Its coupling to matter takes place only indirectly, through the gravitational interaction. By means of a suitable potential, this extra field should reproduce the effect of dark energy. The simplest version of this approach is the famous Brans-Dicke theory \cite{Brans Dicke 1961}. Concerning dark energy, a better version is due to Horndeski \cite{Horndeski 1974}, and it is known as the Fab Four Theory. In it, the scalar field operates as a sum of four terms, each of which plays a crucial role\footnote{To which terms the names John, Paul, George and Ringo, namely the fabulous four, the Beatles, have been given.}. \\
    Among the proposals in this group, there are also the so-called Einstein-\AE ther Theories, which modify General Relativity by adding to the spatio-temporal metric tensor a unit time-like vector field which violates Lorentzian invariance. \\
    Then there are the bimetric theories of gravity (or tensor-tensor theories, like Rosen's \cite{Rosen 1940a, Rosen 1940b} and Drummond's \cite{Drummond 1979} theories), the theories with massive gravity, the TeVeS (tensor-vector-scalar theories) and the STVG (scalar-tensor-vector-gravity), that have MOND theory and bimetric-MOND as non-relativistic limits. \\
    To these we add the Einstein-Cartan-Sciama-Kibble theory \cite{Hehla, Hehlb}, in which spinor fields couple to torsion, a geometrical structure absent in General Relativity. 

    \item \underline{Non-local and higher-derivative theories of gravity}. To this class belong the so-called F(R) theories, built starting by introducing a specific function F of the curvature scalar R. Each choice of F leads to a theory (the most considered are the Carroll-Duvvuri-Trodden-Turner theory \cite{Carroll Duvvuri Trodden Turner 2004} and the Starobinsky one \cite{Starobinsky 1980}). Due to the higher order derivatives they are all plagued by instability problems (Ostrogradsky's \cite{Pons 1988, Stelle 1978, Ostrogradsky 1850}, Frolov's \cite{Frolov 2008}, Dolgov-Kawasaki's \cite{Dolgov Kawasaki 2003, Seifert 2007} instabilities), namely slightly perturbed solutions are subject to large modifications. More general theories are obtained by including all geometrical tensors (the Ricci, Riemann and Weil tensor). Another effective theory is the Ho\v{r}ava-Lifschitz one \cite{Horava 2009a, Horava 2009b}, a non-relativistic theory of gravity that treats space and time as distinct entities. \\
    Then, there are the galileon theory \cite{Nicolis Rattazzi Trincherini 2009} and the multi-galileon theory \cite{Padilla 2010a, Padilla 2010b} that are similar to the scalar-tensor theories. However, in them the scalar field (or fields) appears with derivatives of order higher than the second. \\
    We also mention the ghost condensate theories \cite{Arkani-Hamed 2004}, namely theories that are insensitive to the inclusion of an additive constant to the scalar field, and the non-metric theories \cite{Bengtsson 1991, Bengtsson 1993, Bengtsson 1995, Bengtsson 2007}, in which the gravitational field is no more described by a geometric metric. Finally, in this vein one can insert a theory in which the effects of dark energy are ascribed to curvature corrections determined by interactions between quantum fields and the metric gravitational fields, based on a so-called ultra-strong gravitational principle \cite{Piazza 2009a, Piazza 2009b}.

    \item \underline{Higher-dimensional theories}. Finally, there are theories that assume spacetime to have dimension higher than those we perceive in our daily experience (three spatial dimensions and a temporal one). These may be divided in two categories. The first is the one of the Kaluza-Klein type theories (KK-gravity theories \cite{Freund 1982}) in which one assumes the extra space dimensions to have an extension so small that they turn out to be invisible. However, their effects may manifest themselves as secondary ones in the remaining extended dimensions. The second one assumes, instead, that the extra dimensions are extended, but the whole matter described by the Standard Model of the elementary particles is confined to live on a three-dimensional membrane, so that the effects of the extra dimensions can only be seen as secondary ones (Brane World theories \cite{Akama 1982}). In this group there stand out the Dvali-Gabadadze-Porrati model \cite{Dvali Gabadadze Porrati 2000} and the Randall-Sundrum one \cite{Randall Sundrum 1999a, Randall Sundrum 1999b}. \\
    Another type of multi-dimensional gravity theories are the Einstein-Gauss-Bonnet theories \cite{Zwiebach 1985}. They live in four or five spatial dimensions, where one can add to the theory contributions that couldn't otherwise exist in three space dimensions. Once the model has been constructed, one can bring everything back to the observed dimensions with the KK or Brane World techniques.
\end{enumerate}
There also exist some additional models that lie outside these veins. Among them one finds the Cardassian expansion theory \cite{Freese Lewis 2002}, in which one corrects the FRLW equations by means of extra matter that becomes dominant at large times, thus originating the accelerating expansion, and the theories that postulates the existence of a dark energy spark prior to recombination \cite{Goldstein 2023}. The latter theories are the favourite ones for the solution of the problem of the Hubble tension.

\section{Conclusions: is the Cosmological Constant really constant?}

We have seen that there exist a good many theoretical proposals to explain the accelerated expansion of the universe. But now the word goes again to observations. These should be performed in great numbers and should be very precise, in order to to establish which, among the proposed theoretical explanations (or others, were the present ones to be excluded), would get closer to an explanation of the dynamics of the universe. There are various ongoing projects, each of which requires long observational times and data analysis. We briefly list the most important ones.

\underline{DES} (Dark Energy Survey camera). \\
In 2012, on the Victor M. Blanco telescope at the National Science Foundation's Cerro Tololo Inter American Observatory in Chile, and under the guide of the Fermi National Accelerator Laboratory of the US Department of Energy, a 570 megapixel digital camera was mounted. The latter mapped an area of almost an eighth of the entire sky for a total of 758 nights, stopping taking data in January 2019. The results of the almost final analysis of the data have been presented at a meeting of the American Astronomical Society on January 8, 2024. \\
They are based on the observation of 1499 standardised type Ia supernovae (compared to the 52 used in 1998 by the groups of the 2011 Nobel prize winners S. Perlmutter, B.P. Schmidt and A.G. Riess) and have confirmed with an unprecedented level of accuracy the rate of the accelerated expansion of the universe, for which, as we know, dark energy is responsible. In particular, the standard cosmological model $\Lambda$CDM implies that the dark energy density is constant in space and time. Now, the supernova method constrains three parameters, the Hubble parameter, the matter density of the universe, which decreases in time as the universe expands, and a number called $w$, which indicates whether or not the dark energy density is constant in time, the value of $w$ being $-1$ if this density is indeed constant. The results of the DES analysis found the value $w=-0.80\pm 0.18$. \\
This result, combined with the complementary data of the European Space Agency's Planck satellite is compatible with the value $w=-1$ within the error bars, but does not exclude that the dark energy density might have been varying in time, possibly from almost zero value at very early times, before structures started to form, to its present value of our highly structured universe. Further much more refined additional measurements are necessary to settle this crucial question.

\underline{DESI} (Dark Energy Spectroscopic Instrument). \\
The DESI project has essentially been devised to probe the expansion history of the universe using the imprinting left by the baryon acoustic oscillations (BAO) in the clustering of quasars, galaxies and the intergalactic medium, under the combined effect of gravity and dark energy (see subsection \ref{121}). In particular, it is hoped that DESI will provide some insight into the nature of dark energy, for example whether the latter is constant in space and time or may depend on time. \\
To foster these aims, the scope of the project is to obtain optical spectra for more than 40 million galaxies, quasars and stars, constructing an unprecedented volume of the universe stretching back in time more than 11 billion years into the past. DESI houses 5000 robot actuators adjusting 5000 optical fibres, each one of which can capture the light from a single galaxy or quasar and feed it into one of ten identical spectrographs. The instrument is mounted on the 4 metre Mayall telescope of the Kitt Peak National Observatory in the Sonora desert, in the Indian reservation of the Tohono O'odham, near Tucson, Arizona, and is capable of monitoring more than one hundred thousand luminous sources per night. \\
The first results of cosmic scale have been published in November 2023. They are partly based on the analysis of $261,291$ red galaxies (with redshift $z$ between $0.4$ and $1$), allowing the measurement of the BAO to $5\sigma$ (namely with $99.9994$\% certainty) and angular localisation within $1.7$\%, and partly on the analysis of $109,523$ brilliant galaxies (with redshift between $0.1$ and $0.5$), with a $3\sigma$ precision (corresponding to $99.73$\% certainty) and an angular localisation of $2.6$\%. These correspond to two months of observations and are at the moment considered as a preliminary assessment of the potentiality of the project. Instead, the first conclusive measurements will be based on a cycle of five years.

\underline{PFS} (Prime Focus Spectrograph). \\
It is a spectrographic system endowed with optical fibre actuators and mounted on the Subaru Telescope, an optical and infrared 8 metre telescope located near the top of the inactive Mauna Kea volcano in Hawaii, under the control of the National Astronomical Observatory of Japan. It will allow the spectral analysis of 2400 astronomical targets at a time, in a field of view of $1.3$ degrees in diameter.

\underline{Euclid} (Space Telescope). \\
It is a space telescope developed by the European Space Agency (ESA) whose objective is to better understand dark energy and dark matter by accurately measuring the accelerating expansion of the universe (see subsection \ref{121}). To this aim, it will exploit gravitational lensing, measurement of baryon acoustic oscillations, and measurement of galactic distances by spectroscopy. The telescope travels along a three-dimensional orbit (called halo orbit) around the second Lagrange point (L2) of the Earth-Sun system, which is located at a distance of about $1.5$ million km from Earth along the Earth-Sun axis. At L2, Euclid joins both the Gaia and the James Webb Space Telescope. \\
Euclid is a spacecraft divided into two modules, one containing a $1.2$ metre telescope, a 600 megapixel optical digital camera, and an infrared camera. The second module contains the telescope controls and the thermal and propulsion systems. It is expected to cover one third of the sky, to monitor 10 million galaxies by measuring their redshifts up to a value of $z=2$, corresponding to seeing back $10$ billion years into the past. The total time span of data collection is expected to last six years (up to exhaustion of the fuel). \\
Euclid was launched the $1^{st}$ of July, 2023 and has reached its position on July 30, 2023. It has sent the first preliminary images in November 2023.

\underline{LSST} (Large Synoptic Survey Telescope). \\
It is an $8.4$ metre telescope provided with a $3.2$ gigapixel CCD imaging camera. It is located at the Vera C. Rubin Observatory, on the El Pe\~{n}\'on peak of Cerro Pachon in northern Chile. It is expected to gather imaging data for about 10 years. Tested in May 2022, first light for the engineering camera is expected in August 2024, while system first light is expected in January 2025. First analyses will start in the summer of 2025.

\underline{Nancy Grace Roman Space Telescope}. \\
It is a NASA infrared space telescope scheduled to launch by May 2027. It is provided with a primary $2.4$ metre mirror and a $300.8$ megapixel digital camera. Its multiple aims are: the search of exoplanets by the technique of microlensing; the chronology of the universe and growth of cosmic structures, with special attention to the effects of dark energy; precision verification of general relativity; the measurement of the curvature of spacetime. To probe the effects of dark energy the telescope will employ three independent techniques: baryon acoustic oscillations, observations of distant supernovae, and weak gravitational lensing.

\underline{Extremely Large Telescope (ELT)}. \\
The Extremely Large Telescope is an astronomical observatory that will be the world's largest telescope working in the optical/near--infrared. It is located on top of Cerro Armazones in the Atacama Desert of Northern Chile, and is at present under construction. It will consist of a reflecting telescope with a segmented primary mirror of $39.3m$ of diameter and a secondary mirror of $4.2m$ of diameter. It is projected to have 250 times the light gathering area of the Hubble Space Telescope and to provide images 16 times sharper than those from Hubble. It has a huge light-collecting area of $978m^2$ that is expected to make it able to measure the change in redshift over time of the spectra of distant galaxies, estimated to be about 10 cm/s over a decade, as well as the temperature of the cosmic microwave background radiation and the primordial abundances of light elements. It is planned to take the first images (first light) in 2028. Among the main goals of ELT, there will be the detailed study of planets around other stars, the first galaxies in the universe, and supermassive black holes. In addition, the precision with which ELT will be able to measure the rate of change of the redshift of distant galaxies in the course of time, will help to pin down better the properties of dark energy and, in particular, whether the latter displays a time dependence. ELT will also be able to measure the shape of galaxies' dark matter halos as functions of galaxy mass assembly history and environment. This will help to understand better the nature of dark matter, in particular whether the dark matter halos are made of massive or of extremely light particles, or both.

\

All these missions will be crucial to allow us to select a more precise theoretical model for dark energy.


\end{document}